# Realizing spin-Hamiltonians in nanoscale active photonic lattices

Midya Parto[1,†], William Hayenga[1,†], Alireza Marandi[2], Demetrios N. Christodoulides[1] & Mercedeh Khajavikhan[1,3]*

[1]CREOL, The College of Optics and Photonics,
University of Central Florida, Orlando, FL 32816-2700, USA,
[2]Department of Electrical Engineering, California Institute of Technology, Pasadena, CA 91125, USA
[3]Ming Hsieh Department of Electrical and Computer Engineering, University of Southern California, Los Angeles, California 90089, USA

*Corresponding Author: khajavik@usc.edu
†These authors contributed equally to this work.

**Spin models arise in the microscopic description of magnetic materials, where the macroscopic characteristics are governed by exchange interactions among the constituent magnetic moments. Recently, there has been a growing interest in complex systems with spin Hamiltonians[1–3] – largely due to the rich behaviors exhibited by such interactions at the macroscale. Along these lines, it has been shown that certain classes of optimization problems involving large degrees of freedom can be effectively mapped into classical spin models. In this vein, the respective extremum can be found by identifying the ground state of the associated spin Hamiltonian. Here, we show both theoretically and experimentally, that the cooperative interplay among vectorial electromagnetic modes in coupled metallic nanolasers[4–7] can be utilized as a means to emulate certain types of spin-like systems. The ensuing spin exchange interactions are in general anisotropic, in a way similar to that encountered in magnetic materials involving spin-orbit coupling. For some topologies, we find that these active nanophotonic structures are governed by a classical XY Hamiltonian that exhibits two phases akin to those associated with ferromagnetic (FM) and antiferromagnetic (AF) materials. In addition, we show that in certain configurations, the electromagnetic field distribution can undergo geometrical frustration, depending on the lattice shape and the transverse resonant modes supported by the individual cavity elements. Our results could pave the way towards a new scalable nanophotonic platform to study spin exchange interactions, that can in turn be potentially exploited to investigate more large-scale networks, emulate some magnetic materials, or to address a variety of optimization problems.**

Spin Hmailtonians arise in a ubiquitous manner in nature. Over the years, such models have been a subject of intense investigation in many and diverse areas of science and technology, ranging from condensed matter physics[8] and spintronics[9] to quantum information theory[3]. Of particular interest is geometric frustration that occurs when a certain type of local order, associated with a minimum energy state, cannot extend throughout a system due to geometrical constraints[10]. This effect appears in a variety of physical problems and settings, ranging from residual entropy in water[11] and spin ice[12,13], to orbital exchange in Mott insulators[14] and the emergence of the blue phases in cholesteric liquid crystals[15]. In magnetic materials, frustration is typically associated with a set of highly degenerate ground states of a spin Hamiltonian, which in turn leads to complex macroscopic behaviors such as those observed in spin-liquid or spin-ice phases[16].

In recent years, there has been a number of attempts to cast various computational optimization problems in terms of finding the ground state of a corresponding spin Hamiltonian[1,2]. In this regard, ultracold atomic platforms have been extensively pursued to emulate magnetic-like interactions[17-20]. Alternatively, active optical configurations provide an attractive approach for physically implementing and effectively studying such spin Hamiltonians. In contrast to other methodologies that rely on classical or quantum annealing, optical schemes can quickly converge to the global minimum loss, once gain is introduced. This has incited a flurry of activities in designing photonic "machines" capable of emulating spin exchange Hamiltonians such as, for example, classical Ising or XY Hamiltonians. So far, such active spin systems have been demonstrated using optical parametric oscillators[21-23], polaritonic arrangements[24,25], and degenerate laser cavities[26,27]. Of interest will be to open up new possibilities by introducing additional degrees of freedom through the vectorial nature of the electromagnetic modes in ultracompact spin-like optical resonant structures. These systems can enable addressing large-scale optimization problems in nanoscale integrated platforms, while emulating more complex magnetic materials.

In this Letter, we show how the interactions among resonant vectorial modes in coupled metallic nanolaser arrays can be utilized to emulate certain types of spin Hamiltonians such as for example the classical XY Hamiltonian. By properly designing the array elements, two regimes of exchange coupling can be identified, signifying the occurrence of a ferromagnetic (FM) and an antiferromagnetic (AF) phase (Fig. 1a). Information obtained from the diffraction patterns as well as the winding numbers associated with the vortices formed within the structures, is then used to experimentally characterize these phases. Depending on the vectorial profile of the modes involved, in the AF state, we demonstrate geometrical frustration in various scenarios, in full accord with theoretical predictions. In addition, in certain topologies, this platform can be used to observe anisotropic spin interactions which are known to occur for example in transition-metal compounds and perovskites[28-30]. In what follows, we will outline the mathematical foundation required to explain our results and will present experimental observations corroborating the role of the vectorial nature of light in producing spin-like behaviors in active optical nanocavities – a response that is in sharp contrast with scalar optical settings or oscillator networks[31-33]. Finally, the prospect of expanding such nanolaser platforms in realizing more general spin Hamiltonians is discussed.

To illustrate the spin-like behavior in metallic nanolaser arrays, we consider a circular array of $N$ identical metallic nanodisk lasers, as depicted in the SEM image of Fig. 1b for $N = 6$. Here, the coupling between the nearest-neighbor cavity elements takes place via their overlapping near-fields. The choice of nanolaser as a platform is motivated by: (i) the small cavity size that leads to a large inter-modal spacing (hence individual elements operate robustly in a single mode fashion), (ii) nanolasers are capable of displaying a rich modal diversity because of plasmonic effects. In addition, the metal shell boosts the efficiency of the laser by providing energy confinement, while increasing the Q-factor discrimination between the various modes of the lattice. Finally, the localization of carriers within the small metal cavities prohibits mode competition and spatial hole burning effects as observed in standard multi-moded semiconductor lasers.

In the weak coupling regime, one can obtain the transverse distribution of the resonant electromagnetic fields within the nanodisk $j$ from the corresponding longitudinal component of the electric (magnetic) field of the associated $TM$ ($TE$) mode, i.e. $\psi_j \propto J_n(k_\rho \rho) \cos(n\phi + \phi_j)$. The integer $n$ denotes the azimuthal mode number, while $\phi_j$ is the relative phase with respect to the local coordinates of each site. In such an arrangement, the metallic cladding leads to a mode-dependent dissipation. More specifically, the average power loss for the $TE_{nm}$ and $TM_{nm}$ modes can be expressed as

$$\mathcal{P}_L \propto \begin{cases} \mathcal{P}_1 - \mathcal{P}_{z,TE} \sum_{j=1}^{N} \cos\left(j\frac{2\pi}{N}n + \phi_j\right) \cos\left(-n\pi + j\frac{2\pi}{N}n + \phi_{j+1}\right) \\ \quad + \mathcal{P}_{\phi,TE} \sum_{j=1}^{N} \sin\left(j\frac{2\pi}{N}n + \phi_j\right) \sin\left(-n\pi + j\frac{2\pi}{N}n + \phi_{j+1}\right), \quad TE \\ \\ \mathcal{P}_2 + \mathcal{P}_{\phi,TM} \sum_{j=1}^{N} \cos\left(j\frac{2\pi}{N}n + \phi_j\right) \cos\left(-n\pi + j\frac{2\pi}{N}n + \phi_{j+1}\right), \quad TM. \end{cases} \quad (1)$$

In equation (1), $\mathcal{P}_{1,2}$ represent constant loss terms for the TE, TM modes, while $\mathcal{P}_z$ and $\mathcal{P}_\phi$ depend on the relative strengths of the longitudinal ($H_z$) and transverse ($H_\phi$) magnetic field components, respectively (see Supplementary Part 1). The resulting lasing supermodes supported by this lattice can then be found by minimizing the total loss function $\mathcal{P}_L$ which defines the energy landscape of the system. In order to preserve the discrete symmetry associated with the geometry of the structure, the respective solutions are expected to exhibit a constant discrete rotation $\Delta\phi$ between consecutive cavity elements, i.e. $\phi_{j+1} = \phi_j + \Delta\phi$. Equivalently, the extrema of equation (1) correspond to the minimum energy eigenstates of the following Hamiltonian ($\mathcal{H}$) (see Supplementary Part 1):

$$\begin{aligned} \mathcal{H} &= \mathcal{H}_{XY} + \mathcal{H}_0, \\ \mathcal{H}_{XY} &= \sum_{j=1}^{N} J_{j,j+1} \vec{\sigma}_j \cdot \vec{\sigma}_{j+1}, \\ \mathcal{H}_0 &= \sum_{j=1}^{N} J_{0j,j+1} \cos[\phi_j + \phi_{j+1} + 2j \times 2n\pi/N], \end{aligned} \quad (2)$$

where $\mathcal{H}_{XY}$ is the isotropic XY Hamiltonian describing exchange interactions between the ensuing classical pseudospins defined in each laser cavity as $\vec{\sigma}_j = (\cos\phi_j, \sin\phi_j)$, while $\mathcal{H}_0$ represents an anisotropic Hamiltonian component responsible for lifting the continuous U(1) symmetry within individual cylindrical disks. The corresponding coupling strengths $J$ and $J_0$ for the TE and TM modes are polarization dependent and are respectively given by $J_{TE} = (\mathcal{P}_{\phi,TE} - \mathcal{P}_{z,TE})/2 \times (-1)^n$, $J_{0,TE} = (\mathcal{P}_{\phi,TE} + \mathcal{P}_{z,TE})/2 \times (-1)^{n+1}$ and $J_{0,TM} = J_{TM} = \mathcal{P}_{\phi,TM}/2 \times (-1)^n$. Meanwhile, the second term $\mathcal{H}_0$ in the Hamiltonian can lead to additional minima in the eigenvalue spectrum of the system. Depending on the modes involved, such states with minimal dissipation can coincide with those associated with the original XY Hamiltonian, or may introduce new stable lasing supermodes. In the latter case, the $\mathcal{H}_0$ term acts in a similar way to spin-orbit interactions in certain classes of magnetic materials [28,29] (see Supplementary Part 1).

The Hamiltonian presented in equation (2) can give rise to a variety of field patterns corresponding to ferromagnetic-like (FM) or antiferromagnetic-like (AF) interactions between neighboring pseudospins $\vec{\sigma}_j$. For instance, as shown in Fig. 2a, in a simple two-element arrangement, the resonant $TE_{22}$ mode leads to a negative exchange, $J_{TE} < 0$, that in turn results in an FM-like coupling between the associated pseudospins. On the other hand, for a similar configuration albeit with a slightly different size, once the $TE_{14}$ lasing mode dominates, the underlying coupling becomes positive ($J_{TE} > 0$), as expected from an AF

Hamiltonian (Fig. 2b). When dealing with larger lattices, the AF coupling condition can lead to more complex ground states. In this respect, the competing interactions arising from various nearest neighbor couplings can result in a scenario where the anti-aligned solution is prevented from extending across the entire structure due to geometrical constraints. Consequently, instead of satisfying the minimum energy conditions as dictated by local interactions, the system will eventually minimize its overall energy by settling into a geometrically frustrated state. Perhaps the simplest known example of such an effect is the way three magnetic spins with AF couplings can arrange themselves on a triangle. Figures 2c show two possible degenerate ground states of such a system with opposite winding numbers ($\pm 1$). The geometric frustration in this three-coupled nanolaser configuration is evident in Figs. 2c, d. We note that in this three-element system, the degeneracy between the two eigenstates shown in Figs. 2c, d can be lifted in the presence of $\mathcal{H}_0$, in which case, the vortex mode with winding number $+1$ emerges as the ground state.

In order to experimentally demonstrate the aforementioned FM and AF behaviors in nanolaser networks, we fabricated multiple structures consisting of several coupled elements. Figure 1b shows a scanning electron microscope (SEM) image of an array with $N = 6$ coupled nanodisks before metal deposition. Each nanodisk is cladded with silver, and is separated from its adjacent elements through a metallic silver gap. The gain medium comprises of six InGaAsP quantum wells (with an overall height of 200 $nm$) and is covered by a 10 $nm$ InP layer for protection. The top and bottom sides of each disk are terminated nominally by a 50 $nm$ SiO$_2$ and a 30 $nm$ air plug, respectively (Supplementary Part 2). The nanolaser arrays are characterized using a micro-photoluminescence setup as described in Supplementary Part 3. In order to identify the pertinent cavity lasing modes, we closely study the experimentally observed diffraction patterns for structures of various sizes in terms of their spatial profile, polarization, and wavelength. Along these lines, different states (FM and AF) are promoted by varying the size of the nanodisks involved as well as the lattice configuration. In some cases, laser ambient temperature has been adjusted (78K to room-temperature) in order to match the desired cavity mode with the gain lineshape of the active medium.

Figure 3 presents experimental results demonstrating FM and AF behaviors in arrays involving four nanodisk lasers. For characterizing the FM-like response ($J < 0$), we designed cavity elements having a radius of 575 nm and a 50 nm separation from its nearest neighbors. Our FEM simulations indicate that the individual disks tend to predominately lase in the $TE_{22}$ mode. Meanwhile, from equation (2) one can conclude that in such polygonal arrays comprised of four elements, this same mode can give rise to an FM-like exchange coupling between adjacent pseudospins. The resulting ground state corresponding to this case is illustrated in Figs. 3a-c, along with experimental results, corroborating these predictions. On the other hand, a different design is used in order to observe AF-like interactions. In this case, the nanodisk elements involved in the $N = 4$ array have a radius of 940 nm and are separated from each other by 50 nm. From simulations, this cavity is expected to lase instead in the $TE_{14}$ mode – at a wavelength coinciding with the gain bandwidth of the QWs. Unlike the FM case, here the exchange term is positive ($J > 0$), resulting in an anti-aligned field distribution in the neighboring elements (Figs. 3d-f). The experimental results corresponding to the AF case were obtained at a wavelength of 1415 $nm$. In all cases depicted in Fig. 3, the fields are aligned in such a way so as to enable the system to reach its global minimum in the energy landscape.

We next consider situations where the competing constraints imposed by dissipative interactions among nearest neighbors lead to geometric frustration. Such states occur in AF systems with $J > 0$, in which case the ground state of the optical Hamiltonian no longer follows an anti-aligned field distribution because of the geometry of the lattice configuration itself. To experimentally demonstrate such states, we fabricated two arrays with $N = 3, 5$ elements, where each nanodisk has a radius of 930 $nm$. FEM simulations in this case predict a $TE_{14}$ lasing mode in each element. Figures 4a-h show simulation results of the arrays together with experimentally measured intensity profiles and topological charges associated with the lasing supermodes of these lattices. The associated pseudospins in these cases display a 120° and 144° rotation between consecutive elements, respectively (illustrated in Figs. 4c, g). These results match the geometrically frustrated ground states of the XY Hamiltonian. As mentioned earlier, in some cases, the presence of the second term $\mathcal{H}_0$ in equation (2) poses additional constraints on the ground states of the system. An interesting example in this regard is the case of a hexagonal nanodisk laser arrangement ($N = 6$) with AF interactions, exhibiting geometric frustration due to the anisotropy of $\mathcal{H}_0$ (Fig. 4i). To observe this behavior, we fabricated 850 $nm$ nanodisks, each supporting a $TE_{13}$ lasing mode. In this scenario, the competing interactions described by $\mathcal{H}_{XY}$ and $\mathcal{H}_0$ lead to a frustrated ground state with successive 120° rotations between adjacent pseudospins, as shown in Figs. 4i-k.

The geometrically frustrated states in Fig. 4 represent vortices with a nonzero topological charge. One can map the corresponding orbital angular momenta to a discrete set of spots by monitoring the far-field diffraction patterns after passing through an equilateral triangular aperture[34]. Using this technique, a light beam carrying orbital angular momentum (OAM) with charge $q$, emerges in the far field as a triangular intensity distribution with $|q| + 1$ spots on each side. Moreover, the sign of the associated topological charge can be inferred from the direction of this diffracted triangular pattern (see Supplementary Part 3). Experimental results obtained from such measurements for $N = 3, 5, 6$ lattices are depicted in Figs. 4d, h, l, clearly indicating that in these cases the vortex charge is $l = +1, -2, -2$, respectively. One may improve the characterization precision of the modal profile of the lattice by using near field scanning microscopy.

In order to extend our analysis to larger arrays, we consider a square lattice of FM coupled nanolasers involving 20 × 20 elements (Figs. 5 a-c). Each nanolaser in this structure is designed so as to emit in the $TE_{22}$ lasing mode at $\lambda = 1445\ nm$. The far-field diffraction from this array was experimentally characterized both below and above the lasing threshold. As shown in Fig. 5 d, in the sub-threshold regime, the spontaneously emitted far-field has a Gaussian profile and is unpolarized (Fig. 5 b)– in a way analogous to that anticipated from randomly-oriented pseudospins. As the power of the optical pump is increased, the structure starts to lase and consequently the system settles in the FM ground state of the XY Hamiltonian where the corresponding pseudospins are all aligned (Fig. 5 c). This FM state is corroborated by far-field and polarization measurements – as expected from the emission of aligned vectorial fields in this nanolaser system (Fig. 5 e). In this regard, the pump sets the "temperature" in this platform, as in actual magnetic materials[27,35].

Finally, we investigate disorder effects in large kagome lattices involving AF-coupled nanolasers with 20 × 20 and 40 × 40 unit cells (a total of ~1200 and ~5000 coupled nanodisk lasers, respectively). Figure 6a shows the pseudospins of a possible ground state in such a lattice, as obtained from FEM simulations (Fig. 6 b). To probe long-range order effects in our nanolaser platform, we experimentally imaged the far-field diffraction emitted from these arrays. For a smaller 20 × 20 kagome lattice, the far-field diffraction profile exhibits a sharp hexagonal pattern – as expected from a fully ordered lasing supermode in this particular system. However, for larger kagome lattices (40 × 40 unit cells), we observed a noticeable blurring in the associated far-field diffraction peaks (Fig. 6 d). This observation signifies an absence of long-range order in the associated pseudospins, akin to that encountered in Heisenberg kagome antiferromagnets[26,36,37].

In conclusion, we have shown for the first time that spin exchange Hamiltonians can be realized by exploiting the dissipative cooperative interplay among vectorial electromagnetic modes. The resulting exchange among the corresponding pseudospins is in general anisotropic, akin to spin-orbit coupling in magnetic materials. For certain topologies, the emerging lasing mode is identical to that associated with the ground state of the classical XY Hamiltonian. Using coupled metallic nanolaser lattices, both ferromagnetic and anti-ferromagnetic phases have been demonstrated experimentally, while in the AF regime, geometric frustration has been realized in full accord with theoretical predictions. In general, the proposed optical spin arrangement can be judiciously tailored so as to describe a broader class of spin Hamiltonians with arbitrary exchange couplings $J_{ij}$ (Supplementary Part 5). In principle, it is possible to extend the use of this platform for implementing an Ising Hamiltonian by modifying the lasing mode in individual cavity elements. Moreover, the anisotropic exchange term $\mathcal{H}_0$ that tends to pose a limitation in our system in terms of realizing a pure XY Hamiltonian, can be judiciously neutralized by employing $TE_{0m}$ and $TM_{0m}$ resonant modes. Finally, by providing an integrated solution for implementing photonic spin machines, this platform could pave the way towards emulating more complex networks and material systems wherein a wide class of optimization problems and phase transition phenomena can be studied.

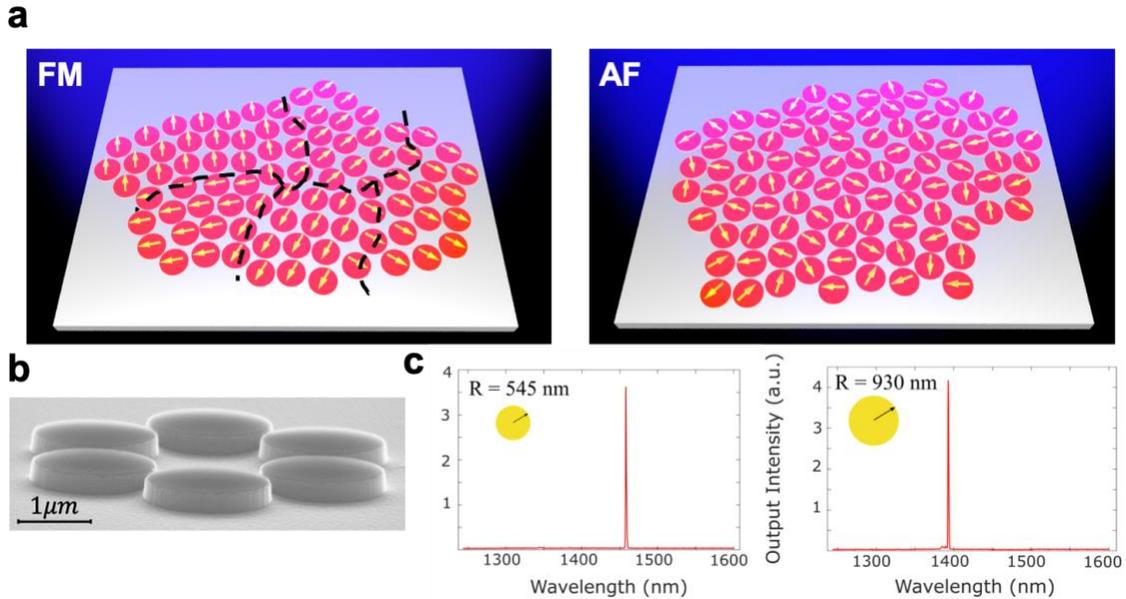

**Fig. 1| Spin-like behavior in coupled metallic nanolaser arrays. a,** Different regimes in spin systems with ferromagnetic (FM) and anti-ferromagnetic (AF) exchange interactions. **b,** SEM image of an array of six coupled active nanodisks (before silver deposition) used in this study. **c,** Measured emitted spectra from single nanodisk lasers with different radii used to emulate FM (left) or AF (right) magnetic spins.

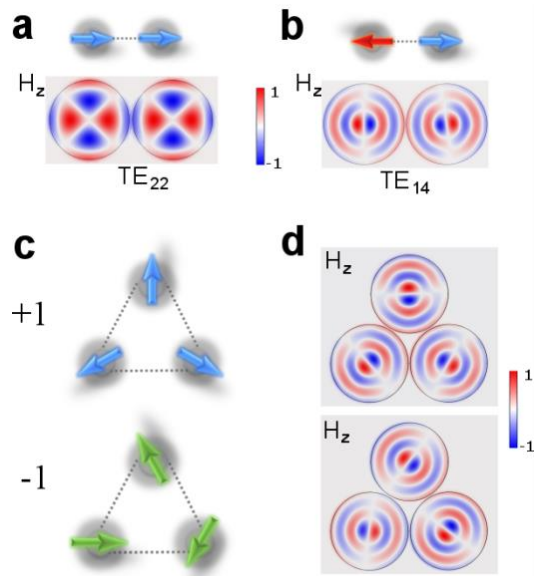

**Fig. 2| Lasing supermodes in coupled nanolasers and their corresponding pseudospins when arranged in simple geometric configurations. a,** Ferromagnetic and **b,** anti-ferromagnetic interactions between pseudospins associated with the longitudinal component of the magnetic field in coupled nanodisk lasers. Depending on the size of the individual elements (575 $nm$ in **a**, or 930 $nm$ in **b**), different $TE_{22}$ and $TE_{14}$ electromagnetic modes predominantly lase, leading to an FM and an AF regime of coupling between nanodisk dimers. **c,** Geometrically frustrated ground states of the classical XY Hamiltonian corresponding to vortices with opposite winding numbers +1 (top) and −1 (bottom). **d,** Resonant electromagnetic supermodes in a triangular array of AF-coupled metallic nanodisks associated with the frustrated states of **c**.

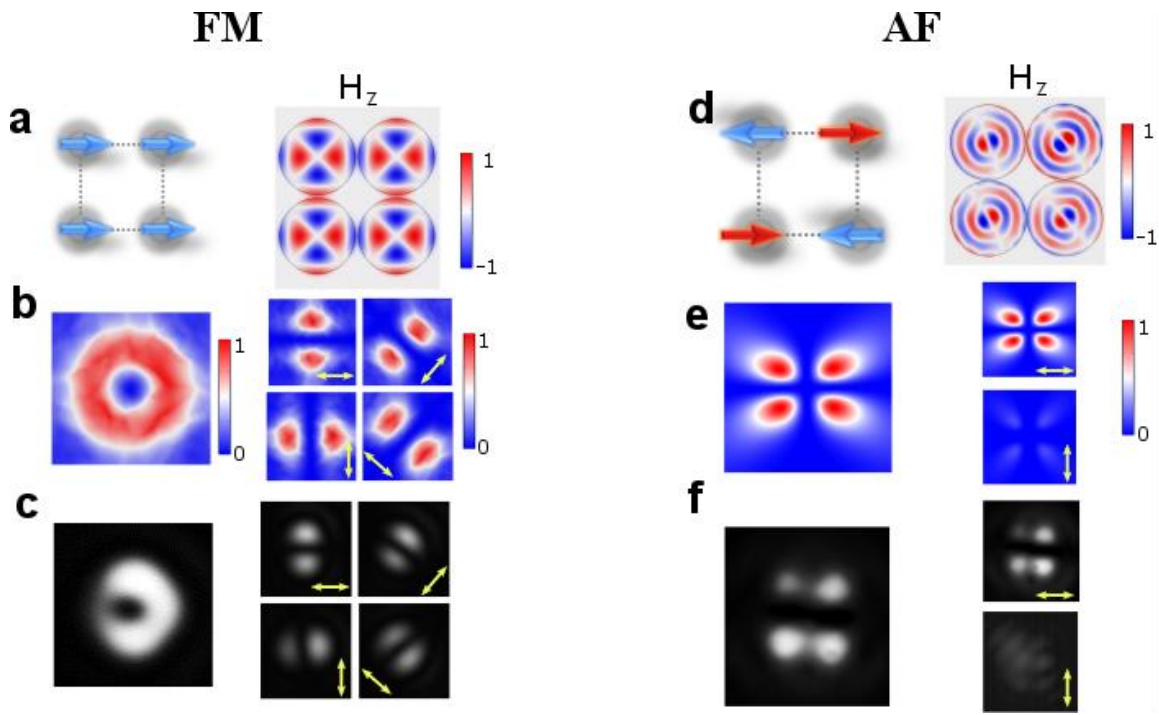

**Fig. 3| FM and AF interactions in four-element coupled nanodisk lasers.** Observation of FM (**a**-**c**) and AF (**d**-**f**) states in different scenarios of coupled metallic nanolasers with various sizes. **a**, Longitudinal magnetic field profiles corresponding to the lasing $TE_{22}$ mode in four coupled nanodisks of radius 575 $nm$. **b**, Theoretically predicted and **c**, experimentally measured optical field intensities and polarization characteristics of the light emitted by such an array. **d-f,** Similar results for the AF case where nanodisks having radii of 940 $nm$ are employed in the same square geometry. The yellow arrows indicate the direction of the linear polarizer. No geometric frustration is observed in either the FM or the AF regimes.

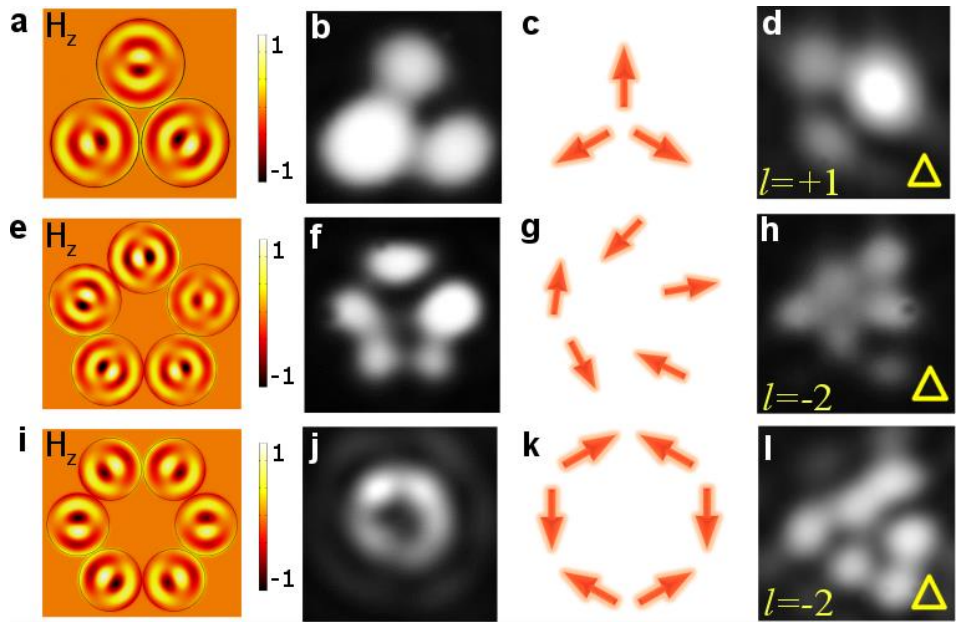

**Fig. 4| Frustrated states in spin-like lasing fields emerging from nanolaser arrays. a,** Simulated lasing profile, **b,** experimentally measured optical intensities, and **c,** associated pseudospin configuration for a geometrically frustrated lasing supermode emitted from an $N = 3$ nanolaser array with an AF-type interaction. In this case each nanodisk has a radius of $930\ nm$, leading to a $TE_{14}$ cavity mode. **d,** Optical intensity pattern obtained after diffraction from a triangular aperture, indicating a topological charge of $l = +1$, as expected from the pseudospin arrangement of **c**. **e-h & i-l** present similar results for $N = 5$ and $N = 6$ nanolaser arrays involving elements with radii $930\ nm$ and $850\ nm$ (TE13 mode), respectively. Note that for these structures the triangular diffraction pattern indicates an $l = -2$ topological charge (**h & l**). The triangular diffraction measurements were all performed by incorporating a $\lambda/4$ waveplate before the aperture followed by a linear polarizer so as to filter for the right-hand circularly polarized component.

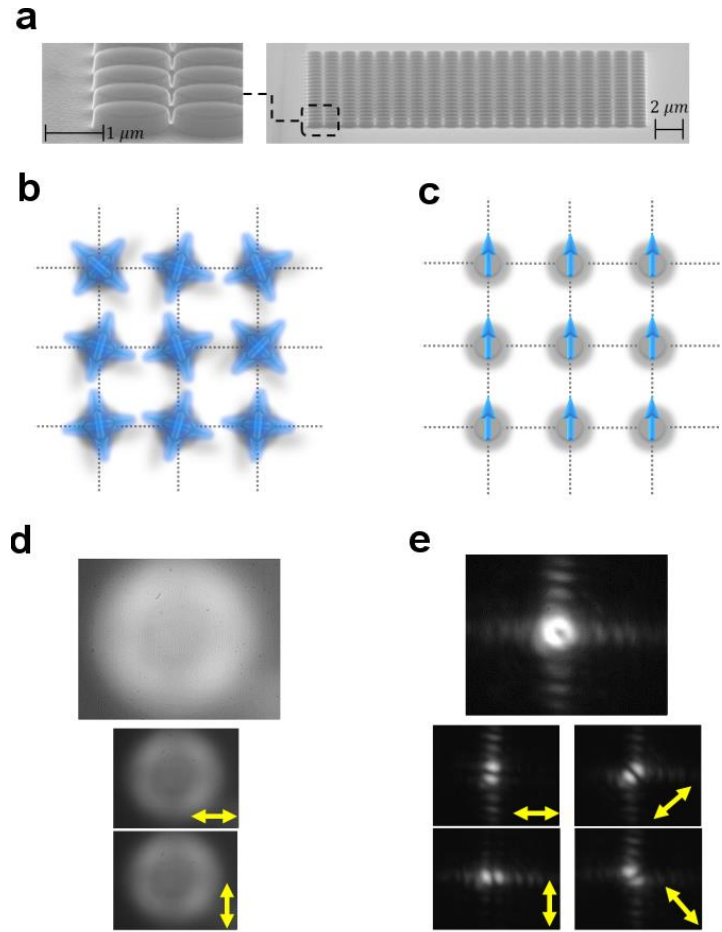

**Fig. 5| Square lattice of 20 × 20 nanolasers exhibiting FM spin-like behavior. a,** SEM image of a 20x20 square lattice of coupled active nanodisks used in this study. **b,** Below the lasing threshold, the orientation of the pseudospins associated with the electromagnetic $TE_{22}$ mode within each laser element are randomly fluctuating. **c,** Once the pump exceeds the threshold, the array starts to lase in an FM state, with the pseudospins aligned in the same direction. **d & e,** Experimentally measured optical intensity patterns and polarization characteristics of the light emitted by this square lattice below (spontaneous emission) and above the lasing threshold, respectively. In the lasing regime, the far-field from this nanolaser square lattice clearly indicates in-phase coherent emission in the $TE_{22}$ mode (ferromagnetic state). The yellow arrows indicate the orientation of the linear polarizer.

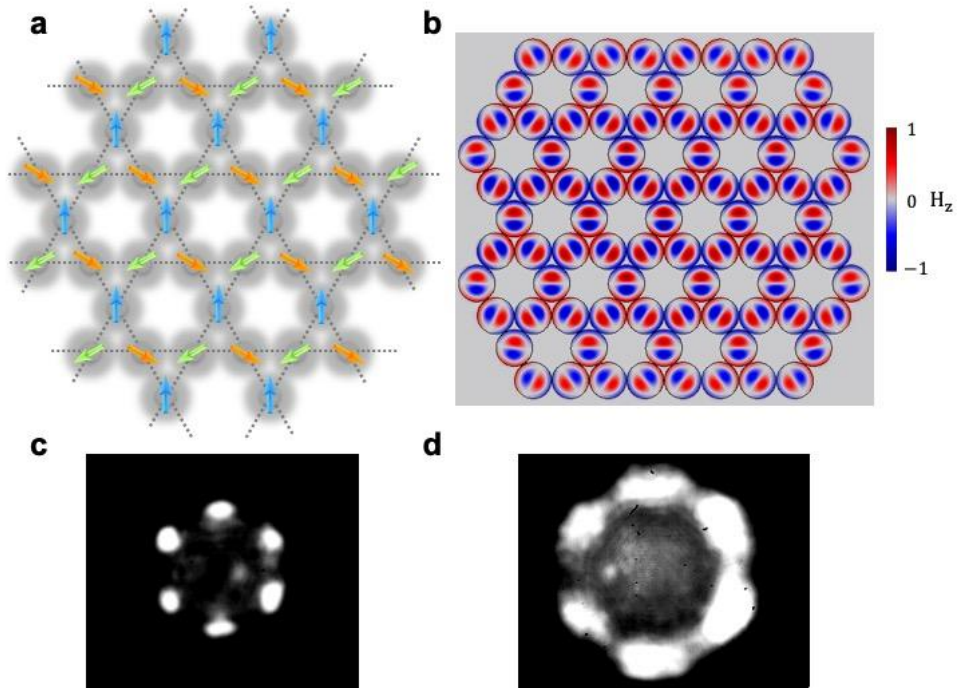

**Fig. 6| Experimnetal Observation of disorder effects in a kagome antiferromagnetic lattice of nanolasers.**
**a**, The pseudospin arrangement of a possible ground state in a kagome lattice. **b**, Simulation results of the lasing supermode corresponding to **a**. Measured far-field diffraction patterns from a **c,** 20 × 20 and **d,** 40 × 40 unit-cell nanolaser kagome lattice. As opposed to **c**, the diffraction spots are now blurred in **d**.

REFERENCES

1. Cuevas, G. D. las & Cubitt, T. S. Simple universal models capture all classical spin physics. *Science* **351**, 1180–1183 (2016).

2. Kirkpatrick, S., Gelatt, C. D. & Vecchi, M. P. Optimization by Simulated Annealing. *Science* **220**, 671–680 (1983).

3. Johnson, M. W. *et al.* Quantum annealing with manufactured spins. *Nature* **473**, 194–198 (2011).

4. Hill, M. T. *et al.* Lasing in metallic-coated nanocavities. *Nature Photonics* **1**, 589–594 (2007).

5. Noginov, M. A. *et al.* Demonstration of a spaser-based nanolaser. *Nature* **460**, 1110–1112 (2009).

6. Oulton, R. F. *et al.* Plasmon lasers at deep subwavelength scale. *Nature* **461**, 629–632 (2009).

7. Khajavikhan, M. *et al.* Thresholdless nanoscale coaxial lasers. *Nature* **482**, 204–207 (2012).

8. Van Vleck, J. H. *The Theory Of Electric And Magnetic Susceptibilities*. (Oxford At The Clarendon Press, 1932).

9. Wolf, S. A. *et al.* Spintronics: A Spin-Based Electronics Vision for the Future. *Science* **294**, 1488–1495 (2001).

10. Sadoc, J.-F. & Mosseri, R. Geometrical Frustration by Jean-François Sadoc. *Cambridge Core* (1999). doi:10.1017/CBO9780511599934

11. Pauling, L. The Structure and Entropy of Ice and of Other Crystals with Some Randomness of Atomic Arrangement. *J. Am. Chem. Soc.* **57**, 2680–2684 (1935).

12. Ramirez, A. P., Hayashi, A., Cava, R. J., Siddharthan, R. & Shastry, B. S. Zero-point entropy in 'spin ice'. *Nature* **399**, 333–335 (1999).

13. Harris, M. J., Bramwell, S. T., McMorrow, D. F., Zeiske, T. & Godfrey, K. W. Geometrical Frustration in the Ferromagnetic Pyrochlore Ho2Ti2O7. *Phys. Rev. Lett.* **79**, 2554–2557 (1997).

14. Wu, C. Orbital Ordering and Frustration of p-Band Mott Insulators. *Phys. Rev. Lett.* **100**, 200406 (2008).

15. Wright, D. C. & Mermin, N. D. Crystalline liquids: the blue phases. *Rev. Mod. Phys.* **61**, 385–432 (1989).

16. Balents, L. Spin liquids in frustrated magnets. *Nature* **464**, 199–208 (2010).

17. Bloch, I., Dalibard, J. & Nascimbène, S. Quantum simulations with ultracold quantum gases. *Nat Phys* **8**, 267–276 (2012).

18. Struck, J. *et al.* Quantum Simulation of Frustrated Classical Magnetism in Triangular Optical Lattices. *Science* **333**, 996–999 (2011).

19. Trotzky, S. *et al.* Time-Resolved Observation and Control of Superexchange Interactions with Ultracold Atoms in Optical Lattices. *Science* **319**, 295–299 (2008).

20. Struck, J. *et al.* Engineering Ising-XY spin-models in a triangular lattice using tunable artificial gauge fields. *Nature Physics* **9**, 738–743 (2013).

21. Marandi, A., Wang, Z., Takata, K., Byer, R. L. & Yamamoto, Y. Network of time-multiplexed optical parametric oscillators as a coherent Ising machine. *Nature Photonics* **8**, 937–942 (2014).

22. McMahon, P. L. *et al.* A fully-programmable 100-spin coherent Ising machine with all-to-all connections. *Science* aah5178 (2016). doi:10.1126/science.aah5178

23. Takeda, Y. *et al.* Boltzmann sampling for an XY model using a non-degenerate optical parametric oscillator network. *Quantum Sci. Technol.* **3**, 014004 (2018).

24. Berloff, N. G. *et al.* Realizing the classical *XY* Hamiltonian in polariton simulators. *Nature Materials* **16**, 1120–1126 (2017).

25. Lagoudakis, P. G. & Berloff, N. G. A polariton graph simulator. *New J. Phys.* **19**, 125008 (2017).

26. Nixon, M., Ronen, E., Friesem, A. A. & Davidson, N. Observing Geometric Frustration with Thousands of Coupled Lasers. *Phys. Rev. Lett.* **110**, 184102 (2013).

27. Pal, V., Tradonsky, C., Chriki, R., Friesem, A. A. & Davidson, N. Observing Dissipative Topological Defects with Coupled Lasers. *Phys. Rev. Lett.* **119**, 013902 (2017).

28. Khomskii, D. I. Transition Metal Compounds by Daniel I. Khomskii. *Cambridge Core* (2014). doi:10.1017/CBO9781139096782

29. Feng, J. S. & Xiang, H. J. Anisotropic symmetric exchange as a new mechanism for multiferroicity. *Phys. Rev. B* **93**, 174416 (2016).

30. Vleck, J. H. V. Recent developments in the theory of antiferromagnetism. *J. Phys. Radium* **12**, 262–274 (1951).

31. Strogatz, S. H. Exploring complex networks. *Nature* **410**, 268–276 (2001).

32. Ariaratnam, J. T. & Strogatz, S. H. Phase Diagram for the Winfree Model of Coupled Nonlinear Oscillators. *Phys. Rev. Lett.* **86**, 4278–4281 (2001).

33. Soriano, M. C., García-Ojalvo, J., Mirasso, C. R. & Fischer, I. Complex photonics: Dynamics and applications of delay-coupled semiconductors lasers. *Rev. Mod. Phys.* **85**, 421–470 (2013).

34. Hickmann, J. M., Fonseca, E. J. S., Soares, W. C. & Chávez-Cerda, S. Unveiling a Truncated Optical Lattice Associated with a Triangular Aperture Using Light's Orbital Angular Momentum. *Phys. Rev. Lett.* **105**, 053904 (2010).

35. Inagaki, T. *et al.* Large-scale Ising spin network based on degenerate optical parametric oscillators. *Nature Photonics* **10**, 415–419 (2016).

36. Moessner, R. & Chalker, J. T. Low-temperature properties of classical geometrically frustrated antiferromagnets. *Phys. Rev. B* **58**, 12049–12062 (1998).

37. Zhitomirsky, M. E. Octupolar ordering of classical kagome antiferromagnets in two and three dimensions. *Phys. Rev. B* **78**, 094423 (2008).

**ACKNOWLEDGMENT**
The authors gratefully acknowledge the financial support from Army Research Office (ARO) (W911NF-16-1-0013, W911NF-17-1-0481), DARPA (D18AP00058, HR00111820042, HR00111820038), National Science Foundation (ECCS 1454531, DMR-1420620, ECCS 1757025, CBET 1805200), Office of Naval Research (N0001416-1- 2640, N00014-18-1-2347), Air Force Office of Scientific Research (FA9550-14-1- 0037), and U.S.-Israel Binational Science Foundation (BSF) (2016381).

**Author contributions:**
All authors contributed equally.

**Competing financial interests:**
The authors declare no competing financial interests.

**Data Availability.** The datasets generated during and/or analyzed during the current study are available from the corresponding author on reasonable request.

## Methods

**Fabrication.** The metallic nanolaser lattices are fabricated on III-V semiconductor platform. The gain material consists of six InGaAsP quantum wells with an overall height of 200 $nm$ grown on an InP substrate. The quantum wells are covered by a 10 $nm$ thick InP over-layer for protection. Hydrogen silsesquioxane (HSQ) solution in methyl isobutyl ketone (MIBK) is used as a negative tone electron beam resist. The lattices are patterned with electron beam lithography, where the exposed HSQ serves as a mask for the subsequent reactive ion etching process. A mixture of $H_2$:$CH_4$:Ar gas chemistry is used with a ratio of 40:6:15 sccm, RIE power of 150 $W$, and ICP power of 150 $W$ at a chamber pressure of 35 $mT$. The wafer is then cleaned with oxygen plasma to remove organic contaminations and polymers that form during the dry etching process. A 1000 $nm$ layer of silver is next deposited onto the lattices by means of electron beam evaporation. SU-8 photoresist is used to bond the wafer to a glass substrate for mechanical support. Lastly, the remaining InP substrate is completely removed by wet etching in hydrochloric acid.

**Experimental arrangement.** The metallic nanolaser lattices is pumped by a pulsed fiber laser operating at a wavelength 1064 $nm$ (15 $ns$ pulse width, 290 $kHz$ repetition rate). The pump beam is focused onto the sample with a 50x objective, this objective in turn also collects the emission from the sample. Light is then either directed to a linear array detector for spectral measurements or to an IR camera for modal profile observation. A triangular aperture is inserted at the back focal plane of the lens before the IR camera for topological charge measurements.

**Code availability.** The codes associated with this manuscript are available from the corresponding author on reasonable request.

# SUPPLEMENTARY

## Part 1. Derivation of the energy landscape function

In order to obtain the resonant modes supported by the nanodisks, we consider the two general sets of transverse electric ($TE$) and transverse magnetic ($TM$) modes. Using separation of variables, and after solving the Helmholtz equation in cylindrical coordinates, one finds the field components for $TE_{nm}$ modes in a nanodisk at site $j$ as:

$$E_{\rho,j} \propto \frac{n}{\rho} J_n(k_\rho \rho) \sin(n\phi + \phi_j),$$

$$E_{\phi,j} \propto \frac{k_\rho}{2} [J_{n-1}(k_\rho \rho) - J_{n+1}(k_\rho \rho)] \cos(n\phi + \phi_j)$$

$$E_{z,j} = 0$$

$$H_{\rho,j} \propto \frac{-\sqrt{k^2 - k_\rho^2} \times k_\rho}{2\omega\mu_0} [J_{n-1}(k_\rho \rho) - J_{n+1}(k_\rho \rho)] \cos(n\phi + \phi_j) \qquad \text{(S-1)}$$

$$H_{\phi,j} \propto \frac{n\sqrt{k^2 - k_\rho^2}}{\omega\mu_0 \rho} J_n(k_\rho \rho) \sin(n\phi + \phi_j)$$

$$H_{z,j} \propto \frac{ik_\rho^2}{\omega\mu_0} J_n(k_\rho \rho) \cos(n\phi + \phi_j).$$

On the other hand, the expression for the total dissipated electromagnetic power due to the metallic walls can be obtained from the surface integrals

$$\begin{aligned}
\mathcal{P}_{L,T} &\propto \sum_{j=1}^{N} \int_{S_j} \left[ |H_{\phi,j}|^2 + |H_{z,j}|^2 \right] ds \\
&- \Delta S \sum_{j=1}^{N} \left\{ |H_{z,j}|^2 + |H_{z,j+1}|^2 - |H_{z,j+1} - H_{z,j}|^2 \right\} \\
&+ \left\{ |H_{\phi,j}|^2 + |H_{\phi,j+1}|^2 - |H_{\phi,j+1} + H_{\phi,j}|^2 \right\},
\end{aligned} \qquad \text{(S-2)}$$

where $S_j$ is the cylindrical surface of the $j$th nanodisk, while $\Delta S$ is the effective common area between nearby nanodisks. From equations (S-1), one can further simplify equation (S-2) as follows

$$\begin{aligned}
\mathcal{P}_{L,TE} &\propto \mathcal{P}_1 - \mathcal{P}_{z,TE} \sum_{j=1}^{N} \cos\left(j\frac{2\pi}{N}n + \phi_j\right) \cos\left(-n\pi + j\frac{2\pi}{N}n + \phi_{j+1}\right) \\
&+ \mathcal{P}_{\phi,TE} \sum_{j=1}^{N} \sin\left(j\frac{2\pi}{N}n + \phi_j\right) \sin\left(-n\pi + j\frac{2\pi}{N}n + \phi_{j+1}\right)
\end{aligned} \qquad \text{(S-3)}$$

where $\mathcal{P}_{z,TE} = 2\Delta S \frac{k_\rho^4}{\omega^2 \mu_0^2} J_n^2(k_\rho a)$, $\mathcal{P}_{\phi,TE} = 2\Delta S \frac{n^2(k^2 - k_\rho^2)}{\omega^2 \mu_0^2 a^2} J_n^2(k_\rho a)$, and $a$ is the radius of the nanodisks. Equation (S-3) can be rewritten as

$$\mathcal{P}_{L,TE} \propto \mathcal{P}_1 + (-1)^{n+1} \sum_{j=1}^{N} \left\{ \frac{\mathcal{P}_{z,TE} - \mathcal{P}_{\phi,TE}}{2} \cos[\phi_{j+1} - \phi_j] \right. \tag{S-4}$$
$$\left. + \frac{\mathcal{P}_{z,TE} + \mathcal{P}_{\phi,TE}}{2} \cos\left[\phi_{j+1} + \phi_j + 2j \times \frac{2n\pi}{N}\right] \right\}.$$

Similarly, for $TM_{nm}$ modes, the fields within cavities are given according to

$$E_{\rho,j} \propto \frac{-\sqrt{k^2 - k_\rho^2} \times k_\rho}{2\omega\epsilon_0} [J_{n-1}(k_\rho\rho) - J_{n+1}(k_\rho\rho)] \cos(n\phi + \phi_j),$$

$$E_{\phi,j} \propto \frac{n\sqrt{k^2 - k_\rho^2}}{\omega\epsilon_0 \rho} J_n(k_\rho\rho) \sin(n\phi + \phi_j)$$

$$E_{z,j} \propto \frac{ik_\rho^2}{\omega\epsilon_0} J_n(k_\rho\rho) \cos(n\phi + \phi_j) \tag{S-5}$$

$$H_{\rho,j} \propto -\frac{n}{\rho} J_n(k_\rho\rho) \sin(n\phi + \phi_j)$$

$$H_{\phi,j} \propto -\frac{k_\rho}{2} [J_{n-1}(k_\rho\rho) - J_{n+1}(k_\rho\rho)] \cos(n\phi + \phi_j)$$

$$H_{z,j} = 0.$$

Therefore, the total dissipated power can again be obtained in a similar way:

$$\mathcal{P}_{L,TM} \propto \mathcal{P}_2 + \mathcal{P}_{\phi,TM} \sum_{j=1}^{N} \cos\left(j\frac{2\pi}{N}n + \phi_j\right) \cos\left(-n\pi + j\frac{2\pi}{N}n + \phi_{j+1}\right), \tag{S-6}$$

where $\mathcal{P}_{\phi,TM} = \frac{k_\rho^2}{2} \Delta S [J_{n-1}(k_\rho\rho) - J_{n+1}(k_\rho\rho)]^2$. Using this, it is straightforward to show that

$$\mathcal{P}_{L,TM} \propto \mathcal{P}_2 + (-1)^n \mathcal{P}_{\phi,TM} \sum_{j=1}^{N} \left\{ \frac{1}{2} \cos[\phi_{j+1} - \phi_j] + \frac{1}{2} \cos\left[\phi_{j+1} + \phi_j + 2j \times \frac{2n\pi}{N}\right] \right\}. \tag{S-7}$$

Equations (1) and (2) in the main text can be directly extracted from (S-3), (S-6) and (S-4), (S-7), respectively. These equations provide the energy landscape functions associated with the Hamiltonians of this system.

In order to see how the energy landscape function can represent an anisotropic XY Hamiltonian, one can rewrite equation (S-3) in the following form:

$$\mathcal{P}_{L,TE} \propto \mathcal{P}_1 - \mathcal{P}_{z,TE} \times (-1)^n \sum_{j=1}^{N} \left\{ \left[\cos j\frac{2\pi n}{N} \cos\phi_j - \sin j\frac{2\pi n}{N} \sin\phi_j\right] \times \right.$$
$$\left[\cos j\frac{2\pi n}{N} \cos\phi_{j+1} - \sin j\frac{2\pi n}{N} \sin\phi_{j+1}\right] \right\} + \mathcal{P}_{\phi,TE} \times (-1)^n \sum_{j=1}^{N} \left\{ \left[\sin j\frac{2\pi n}{N} \cos\phi_j + \right.\right.$$
$$\left.\left. \cos j\frac{2\pi n}{N} \sin\phi_j\right] \times \left[\sin j\frac{2\pi n}{N} \cos\phi_{j+1} + \cos j\frac{2\pi n}{N} \sin\phi_{j+1}\right] \right\} = \mathcal{P}_1 - \mathcal{P}_{z,TE} \times$$
$$(-1)^n \sum_{j=1}^{N} \left\{ \cos^2 \frac{2\pi n j}{N} \cos\phi_j \times \cos\phi_{j+1} + \sin^2 \frac{2\pi n j}{N} \sin\phi_j \times \sin\phi_{j+1} - \frac{1}{2} \sin \frac{4\pi n j}{N} \sin(\phi_j + \right. \tag{S-8}$$
$$\left. \phi_{j+1}) \right\} + \mathcal{P}_{\phi,TE} \times (-1)^n \sum_{j=1}^{N} \left\{ \sin^2 \frac{2\pi n j}{N} \cos\phi_j \times \cos\phi_{j+1} + \cos^2 \frac{2\pi n j}{N} \sin\phi_j \times \sin\phi_{j+1} + \right.$$
$$\left. \frac{1}{2} \sin \frac{4\pi n j}{N} \sin(\phi_j + \phi_{j+1}) \right\}.$$

After rearranging, one would obtain

$$\mathcal{P}_{L,TE} \propto$$
$$= \mathcal{P}_1 + (-1)^n \sum_{j=1}^{N} \left\{ \left[ \mathcal{P}_{\phi,TE} \sin^2 \frac{2\pi n j}{N} - \mathcal{P}_{z,TE} \cos^2 \frac{2\pi n j}{N} \right] \cos \phi_j \times \cos \phi_{j+1} \right. $$
$$+ \left[ \mathcal{P}_{\phi,TE} \cos^2 \frac{2\pi n j}{N} - \mathcal{P}_{z,TE} \sin^2 \frac{2\pi n j}{N} \right] \sin \phi_j \times \sin \phi_{j+1}$$
$$\left. + \frac{1}{2} \sin \frac{4\pi n j}{N} \left[ \mathcal{P}_{\phi,TE} - \mathcal{P}_{z,TE} \right] \sin(\phi_j + \phi_{j+1}) \right\}. \quad \text{(S-9)}$$

From here, one would obtain

$$\mathcal{H} = \sum_{j=1}^{N} \left\{ J_{x,j} \sigma_{x,j} \sigma_{x,j+1} + J_{y,j} \sigma_{y,j} \sigma_{y,j+1} + \vec{\sigma}_j \cdot \mathbf{\Gamma}_{j,j+1} \cdot \vec{\sigma}_{j+1}^T \right\} = \mathcal{H}_{XY} + \mathcal{H}_0, \quad \text{(S-10)}$$

where $J_{x,j} = (-1)^n \left( \mathcal{P}_{\phi,TE} \sin^2 \frac{2\pi n j}{N} - \mathcal{P}_{z,TE} \cos^2 \frac{2\pi n j}{N} \right)$, $J_{y,j} = (-1)^n \left( \mathcal{P}_{\phi,TE} \cos^2 \frac{2\pi n j}{N} - \mathcal{P}_{z,TE} \sin^2 \frac{2\pi n j}{N} \right)$ and

$$\mathbf{\Gamma}_{j,j+1} = \frac{1}{2} \sin \frac{4\pi n j}{N} (-1)^n [\mathcal{P}_{\phi,TE} - \mathcal{P}_{z,TE}] \begin{bmatrix} 0 & 1 \\ 1 & 0 \end{bmatrix}. \quad \text{(S-11)}$$

It is evident from equation (S-11) that if $4n/N = m$ is an integer, then the equivalent spin Hamiltonian associated with this energy landscape function reduces to the XY Hamiltonian with a lifted U(1) symmetry.
Similar results can be obtained for TM modes.

## Part 2. Fabrication procedure and SEM images

The fabrication steps involved in implementing metallic nanolaser lattices are depicted below in Fig. S1. The wafer (grown by OEpic Inc.) consists of six quantum wells of $In_{x=0.734}Ga_{1-x}As_{y=0.57}P_{1-y}$ (thickness: 10 nm), each sandwiched between two cladding layers of $In_{x=0.56}Ga_{1-x}As_{y=0.938}P_{1-y}$ (thickness: 20 nm), with an overall height of 200 nm, grown on an InP substrate. The quantum wells are covered by a 10 nm thick InP over-layer for protection (Fig. S1a). An XR-1541 hydrogen silsesquioxane (HSQ) solution in methyl isobutyl ketone (MIBK) is used as a negative electron beam resist. The resist is spun onto the wafer, resulting in a thickness of 50 nm (Fig. S1b). The lattices are then patterned by electron beam lithography (Fig. S1c). The wafer is next immersed in tetramethylammonium hydroxide (TMAH) to develop the patterns. The HSQ exposed to the electron beam now remains and serves as a mask for the subsequent reactive ion etching process. To perform the dry etching, a mixture of $H_2$:$CH_4$:Ar gas is used with a ratio of 40:6:15 sccm, RIE power of 150 W, and ICP power of 150 W at a chamber pressure of 35 mT (Fig. S1d). The wafer is then cleaned with oxygen plasma to remove organic contaminations and polymers that form during the dry etching process. After this, a 1000 nm layer of silver is deposited onto the sample using electron beam evaporation at a pressure of $5 \times 10^{-7}$ Torr at a rate of $0.1 \: \dot{A}/s$ for the first 400 nm, at which point the rate is ramped up to $1 \: \dot{A}/s$ (Fig. S1e). SU-8 is then used to bond the silver side to a glass substrate for support (Fig. S1f). Lastly, the sample is wet etched in hydrochloric acid to remove the InP substrate (Fig. S1g). SEM images of all lattices discussed in the main text are provided in Fig. S2, after the intermediate dry etching step.

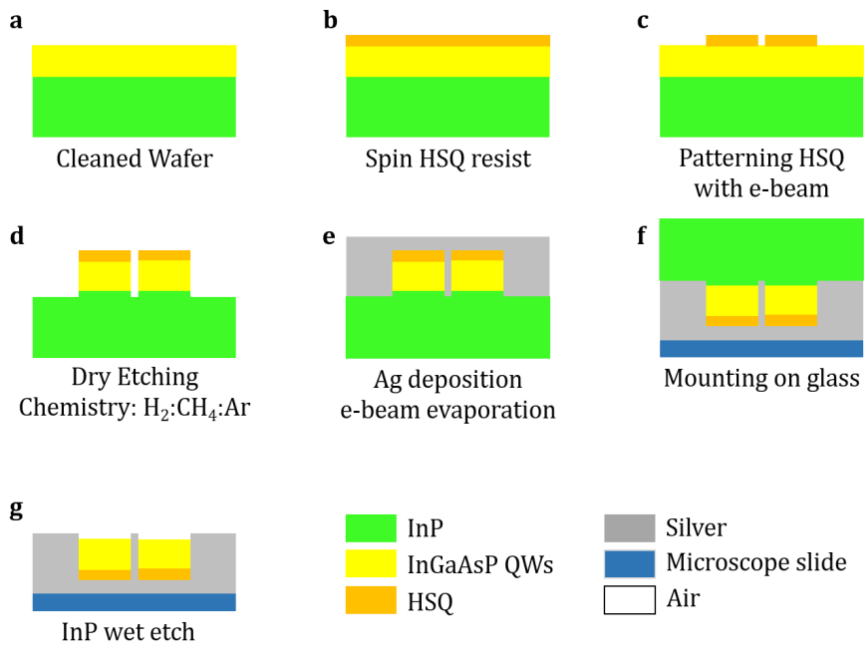

**Fig. S1| Fabrication process of a metallic nanolaser lattice.** This same process is used in other geometries as well. The bottom right corner provides a legend to the materials of the structure. **a,** Cleaned wafer with InGaAsP quantum wells grown on an InP substrate. **b,** A thin layer of negative tone HSQ ebeam resist is spun onto the sample. **c,** The wafer is patterned by ebeam lithography and the resist is developed. **d**, A dry etching process is used to define the lattice. **e,** 1 μm Ag is deposited by means of ebeam evaporation. **f,** The sample is mounted and bonded to a glass microscope slide silver side down with SU-8. **g,** Sample is immersed into HCl to remove the InP substrate.

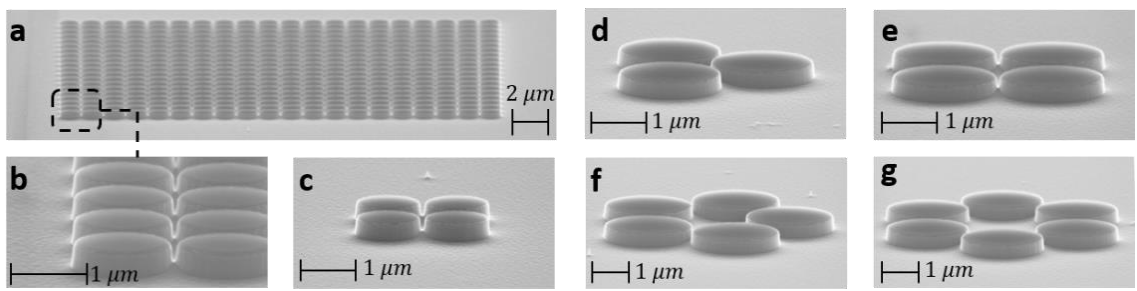

**Fig. S2| SEM images of the lattices.** In all lattices the nanodisks elements are separated by 50 nm. **a,** 20 x 20 array of nanodisks (radius 545 nm). **b,** zoomed in view of the front left corner of the 20 x 20 array. **c,** 4-element array of nanodisks (radius 575 nm). **d, e, & f**, 3-, 4-, & 5-element arrays (radius 930 nm, 940 nm, and 930 nm, respectively). **g,** 6-element array (radius 850 nm).

## Part 3. Characterization setup schematic

A micro-photoluminescence (μ-PL) setup, depicted in Fig. S3, is used to characterize the metallic nanolaser lattices. The lattices are optically pumped by a pulsed laser (duration: 15 ns, repetition rate: 290 kHz) operating at a wavelength of 1064 nm (SPI fiber laser). A beam shaping system is implemented to realize the desired pump profile. In this study, the pump focus spot on the sample has a diameter of 45 *µm*. A 50x microscope objective (NA: 0.42) is used to project the pump beam on the lattice and also serves to collect the emission. For temperature tuning, the sample is inserted into a cryostat (Janis ST-500) and cooled. The surface of the sample is imaged by two cascaded 4-f imaging systems in an IR camera (Xenics Inc.). A broadband ASE source passed through a rotating ground glass is used to illuminate the sample surface for pattern identification. A notch filter is placed in the path of emission to attenuate the pump beam. Output spectra are obtained by a monochromator equipped with an attached InGaAs linear array detector. A powermeter is inserted at the focus of the beam to collect the output power of the laser lattices. A linear polarizer is placed in the setup to observe the polarization resolved intensity distribution. To measure the topological charge of the light emitted by the lattices, a removable equilateral triangular aperture is inserted at the back focal plan of the lens before the IR camera, to facilitate a Fourier-transform. A quarter-wave plate is used to extract the right- and left-handed components of the polarization.

To measure the orbital angular momentum of the emitted light emitted, different techniques can be used based on interference, including self-interference and interference of the OAM beam with a plane wave. In these circumstances, the phase information is assessed by analyzing the fringes. In particular, the formation of a fork-like structure at the center of the vortex is an indication of a topological charge. Here the topological charge measurements are augmented using a different approach based on the relationship between the phase of the light carrying OAM and its diffraction [34]. This latter technique provides an unambiguous measurement of the order and sign of a vortex beam's topological charge. In short, when a *scalar* light beam carrying orbital angular momentum passes through an equilateral triangular aperture, the beam diffracts, hence generating a truncated triangular optical lattice rotated by $\pm 30°$, with respect to the aperture, in the far-field. This lattice then reveals the value of the topological charge ($q$), given by the relationship $|q| = s-1$, where $s$ is the number of spots along each side of the formed triangular lattice. The sign of the charge is determined by the direction that the triangle rotates. For example, in our setup, a triangle pointed right has a positive OAM, while the sign is negative if it points left.

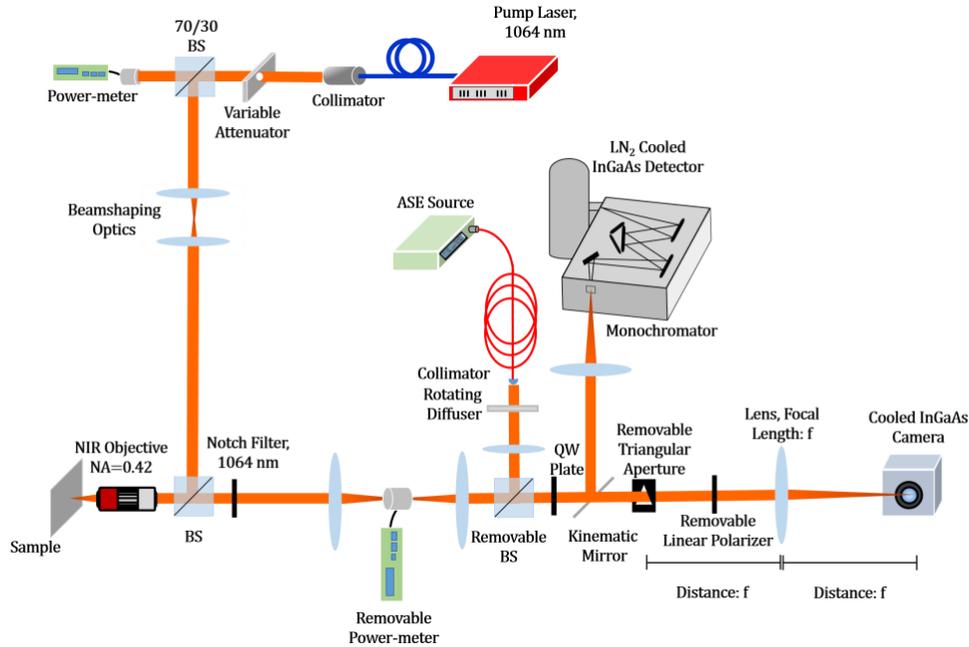

**Fig. S3| Characterization microphotoluminescence setup schematic.** The samples are pumped with a 1064 nm fiber laser and focus onto the sample with a 50x objective (N.A. 0.42). Light is either directed to an InGaAs camera to observe emission patterns, or to a monochromator with an attached InGaAs linear array detector. The emission power is collected at the location of the first focus after the pump notch filter. The triangular aperture used to reveal the OAM of the radiated beams is removable so as to not obstruct imaging the sample surface and to collect proper modal profiles.

To establish the validity of the triangular aperture approach for measuring the topological charge of an optical vortex beam, we compare the results of the triangle technique experimentally to that of a simulated vectorially rotating electric field. In the experiment, we tested a 7-element lattice laser at room temperature (Fig S4 a). As the $TE_{13}$ is a quasi-linear like mode, we simulate this structure by considering six dipoles arranged in a hexagonal ring, emitting radiation with an electric field $\vec{E} \propto -\cos(2\varphi)\hat{x} + \sin(2\varphi)\hat{y}$, where $\varphi$ is the angle of rotation along the periphery. This in turn can be represented in terms of left- ($\hat{L}$) and right-hand circular polarizations ($\hat{R}$)(L,RHCP), $\vec{E} \propto -e^{i2\varphi}\hat{L} - e^{-i2\varphi}\hat{R}$. Figures S4 b, c, & d provide the simulated intensity and the polarization resolved intensity distributions, which are in excellent agreement with the observed modal profiles (Fig. S4, e, f, & g). Figure S4 h shows the result of the simulated diffraction pattern when filtering for the RHCP, and compares it to the experimentally observed diffraction profile (Fig. S4 i). Lastly, the associated light-light (L-L) and spectral evolution curves are provided in Fig. S4, j & k, respectively, clearly showing the threshold characteristic associated with lasing as well as linewidth narrowing.

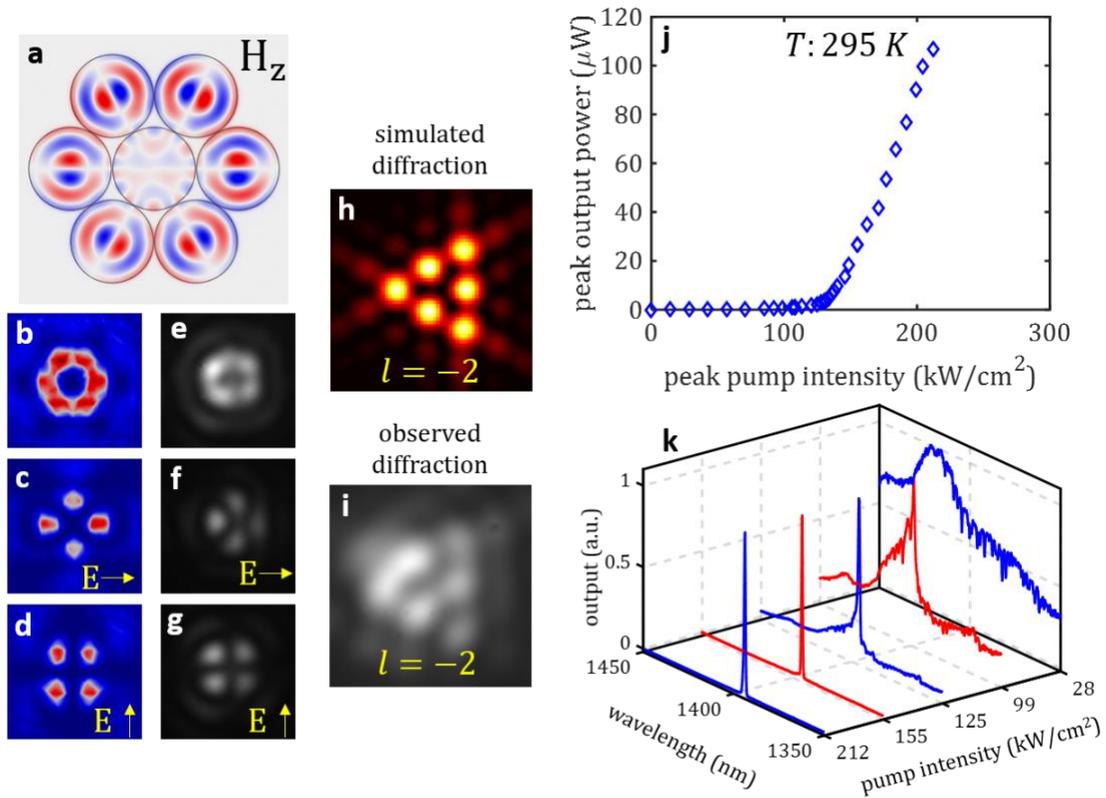

**Fig. S4| Characterization of charge measurement using a 7-element metallic lattice operating at room temperature. a,** Simulated z-component of the magnetic field. **b, c, & d,** The simulated intensity profile and the polarization resolved intensity distributions when filtered for horizontal (**c**) and vertical (**d**) polarizations. **e, f, & g,** The experimentally observed emission pattern and intensity distributions. **h,** The simulated far-field diffraction pattern of the lattice after passing through a triangular aperture. **i,** The experimentally observed profile is in agreement and has $l=-2$. **j,** The light-light curve shows a clear onset in lasing. **k,** The spectral evolution of this laser shows linewidth narrowing and single-mode behavior.

## Part 4. Light-light curves and spectral evolutions

In addition to the emission profile and topological charges provided in the main text, light-light and spectral evolution curves were also collected for these metallic nanolaser lattices. As mentioned before, the ambient temperature of the lasers was tuned in order to promote lasing in certain desired modes. All of the curves shown in this section of the supplementary display a threshold and a linewidth narrowing behaviors with increased pumping, attributes associated with lasing. The reported pump intensities represent the incident light at the sample surface.

Figure S5 provides the L-L and spectral evolutions of the four lattices that orient their fields in an antiferromagnetic-like arrangement. The 3-, 4-, & 5-element structures have individual element radii of 930 nm, 940 nm, and 930 nm, respectively, ($TE_{14}$ mode) and are tuned to a temperature of 225 K (Fig. S5 a-f). On the other hand, the 6-element lattice (nanodisk radius of 850 nm) is left at room temperature to promote lasing in the $TE_{13}$ mode (Fig. S5 g & h).

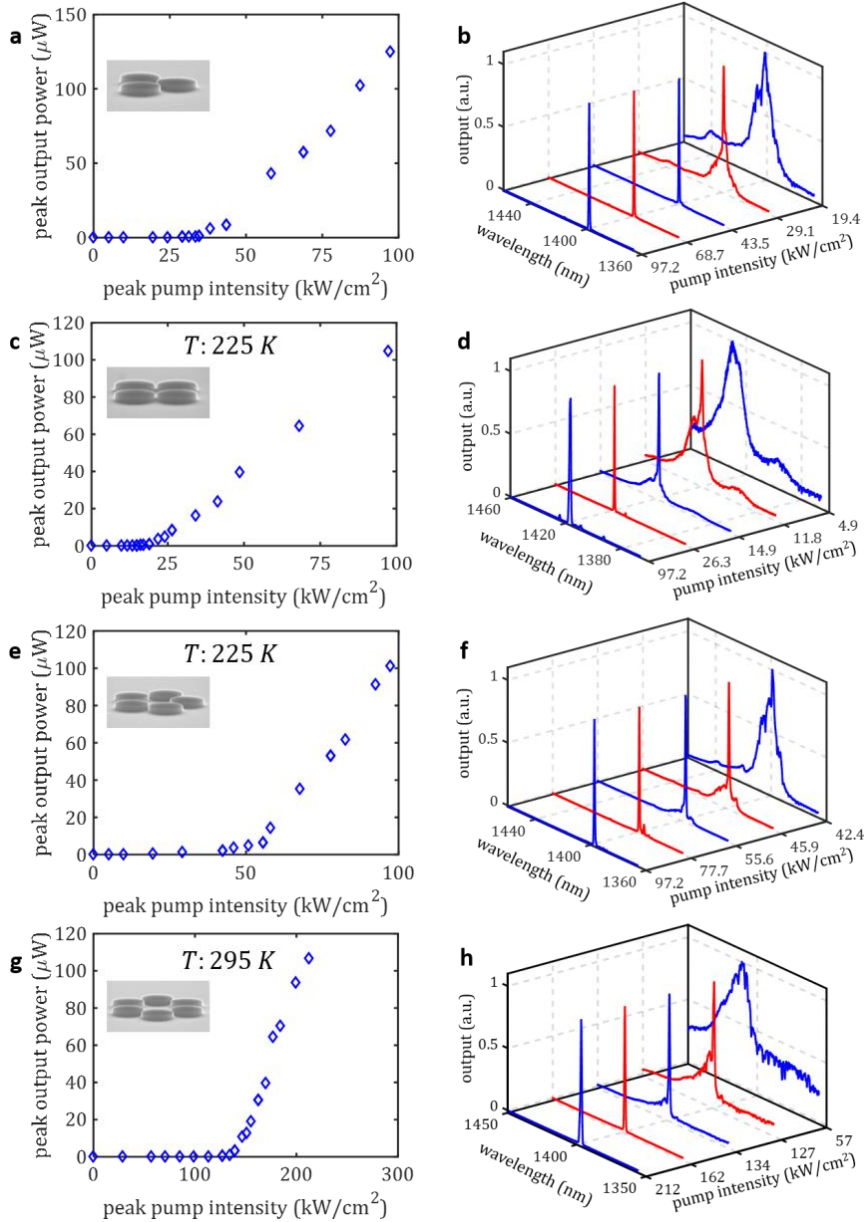

**Fig. S5| Light-light and spectral evolutions curves for the lattices with antiferromagnetic-like coupling.** All the lattices indicate a clear onset of lasing and linewidth narrowing. **a & b, c & d, and e & f,** show the light-light and the spectral evolutions of lattices operating in the $TE_{14}$ mode at temperature of 225 K for the 3-, 4-, & 5-element lattices, respectively. The radii of the nanodisks are 930 nm, 940 nm, and 930 nm, respectively. **g & h** the 6-element lattice supports lasing at room temperature in the $TE_{13}$ mode (850 nm radius elements).

Figure S6 shows the curves for the lattices that orient their fields in a ferromagnetic-like manner. In these lattices all the cavities support the $TE_{22}$ mode. The 4-element arrangement operates at a temperature of 78 K and the characteristics of the L-L and spectral evolution (575 nm nanodisk radius) is provided in Fig. S6 a & b. In the case of the 20 × 20 element array, the nanodisks have a smaller radius of 545 nm and were operated at an ambient temperature of 225 K (Fig. S6 c & d).

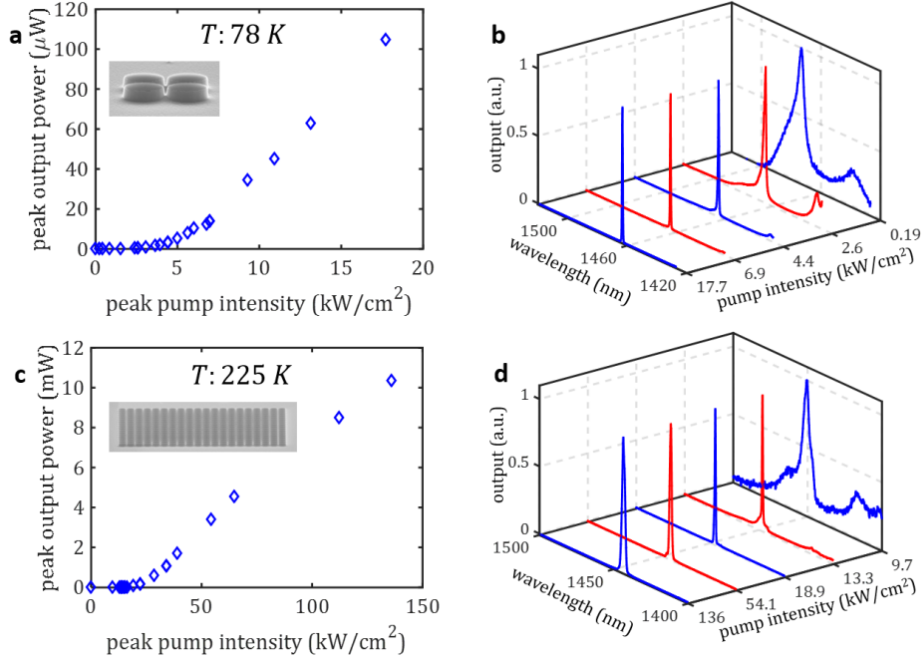

**Fig. S6| Output charactersitics of the 4-element and 20×20-element array displaying ferromagnetic like coupling.** Both lattices operate in the $TE_{22}$ mode, albeit at differing temperatures. **a & b,** the 4-element arrangement is cooled to 78 K and has disks with radii of 575 nm. **c & d**, the 20 × 20-element lattice (radius of 545 nm) generates more than 10 mW peak output power.

## Part 5. Tuning the exchange couplings $J_{ij}$

In order to show the versatility of the coupled nanolaser array platform to realize XY Hamiltonians with various exchange couplings, one may consider varying the metallic gap between nanodisks as a means to adjust the associated exchange couplings in equation (2). To demonstrate this aspect, we study an asymmetric three-element configuration, as depicted schematically in Fig. S7 a. In this configuration, we adjust the relative strengths of the corresponding exchange couplings in the equivalent Hamiltonian of equation (2) such that $J_{12} = J_{13} \approx J_{23}/3$. This is obtained by incorporating a larger gap distance between the top nanodisk and the remaining ones on the bottom of the structure (50 $nm$ versus 25 $nm$). Each of the nanodisks in this case have a radius of 775 $nm$ and emit in the $TE_{13}$ mode. In this case, one expects that the previous 120° arrangement of the pseudospins in an equilateral geometry to be modified towards an anti-aligned pseudospin configuration for the sites located on the bottom of the triangle, as expected from the limiting case of an AF-like coupled dimer geometry. Figure S7 b shows simulated field profile of the associated supermode with the lowest loss in such a geometry, as expected from such asymmetric couplings. Figures S7 c-h display the measured diffracted field intensities and polarization characteristics of light emitted by such a structure (top) together with the associated simulation results (bottom). The coupling between the adjacent elements can be further tuned by depositing additional metallic barriers using focused ion beams.

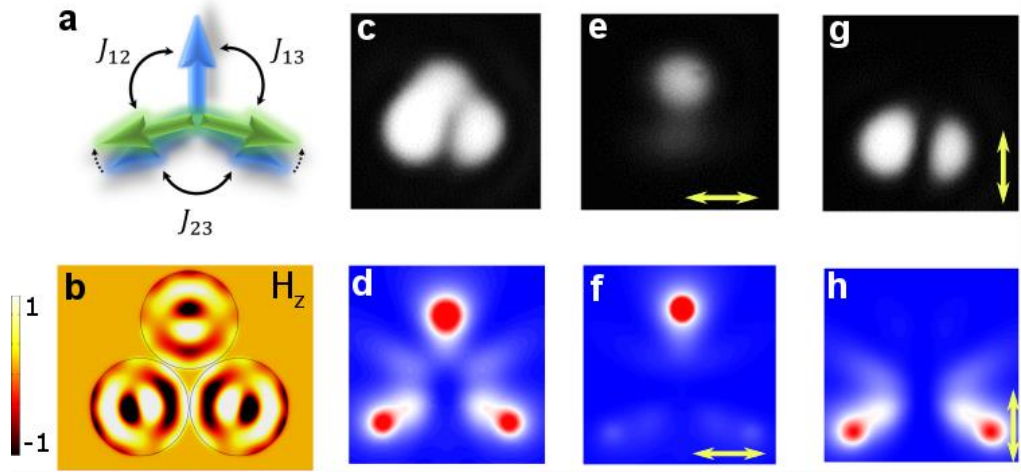

**Fig. S7| Asymmetric triangle geometry. a,** A schematic of three pseudospins arranged on a triangle with asymmetric exchange couplings. In this case, the angles of the bottom pseudospins are expected to change as shown in the figure. **b,** FEM simulation of the lasing supermode in an asymmetric triangular array of nanodisks. Each nanodisk supports a $TE_{13}$ mode. **c-h** Experimental measurements (top) together with theoretically calculated results for diffraction intensities and polarization states of the optical fields emitted by such a nanodisk array. The arrows depict the direction of the linear polarizer.

**Part 6. Polarization measurements for frustrated states**

To further characterize the lasing supermodes in the case of arrays with $N = 3, 5, 6$ nanodisk lasers with an AF-type coupling, we performed polarization measurements in each case and compared the results with those expected from simulations (Fig. S8). These observations further corroborate the results in Fig. 4 of the manuscript.

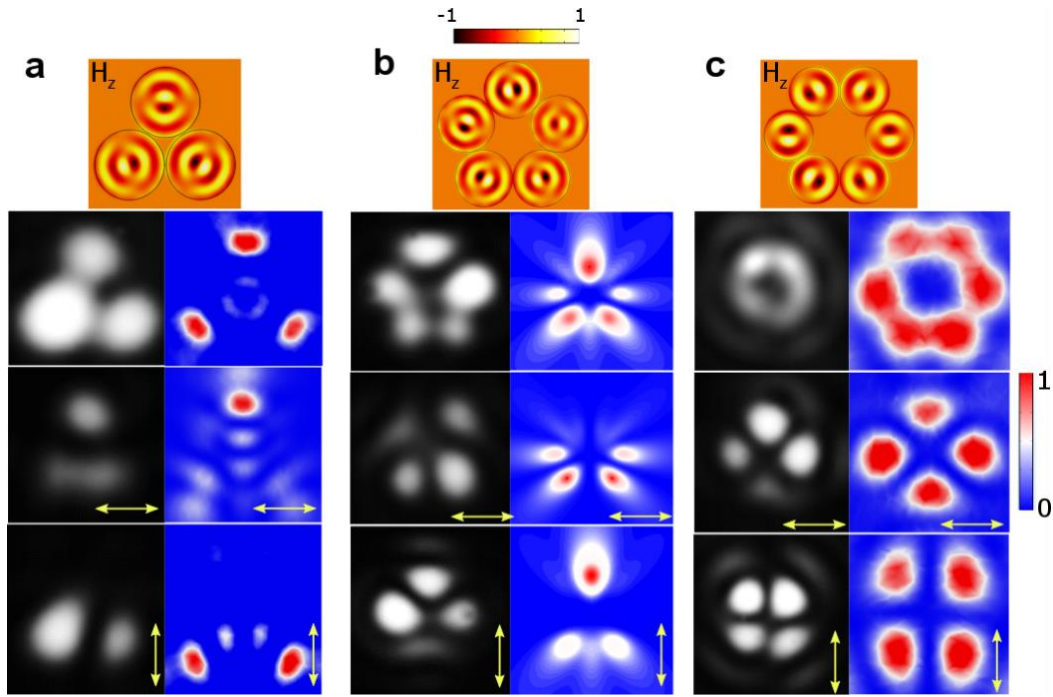

**Fig. S8| Polarization measurements for frustrated states.** Theoretically predicted and experimentally measured polarization profiles for lasing supermodes in arrays of **a,** $N = 3$, **b,** $N = 5$, and **c,** $N = 6$ nanodisk laser arrays. The arrows indicate the direction of linear polarizations.

## Part 7. Characterization of single element nanolasers

To gain a deeper insight into the behavior of various arrays of nanolasers, and to assure the absence of any multimode lasing in the individual elements, we carefully characterized several single element nanolasers, fabricated in various sizes and geometries (nanodisk as well as coaxial). Figure S9 shows simulated resonant modes with highest quality factors of nanodisks with different radii (545 and 930 nm, as used in the FM and AF arrangements throughout this study). In both cases, only one highly confined mode coincides with the gain bandwidth of the quantum well system. Experimental results obtained from these elements are also provided in the same figure, clearly showing single mode lasing based their emitted spectra. Figure S10 further shows additional measured spectra from other nanolaser single elements. In all the cases studied, consistent single-mode lasing was observed.

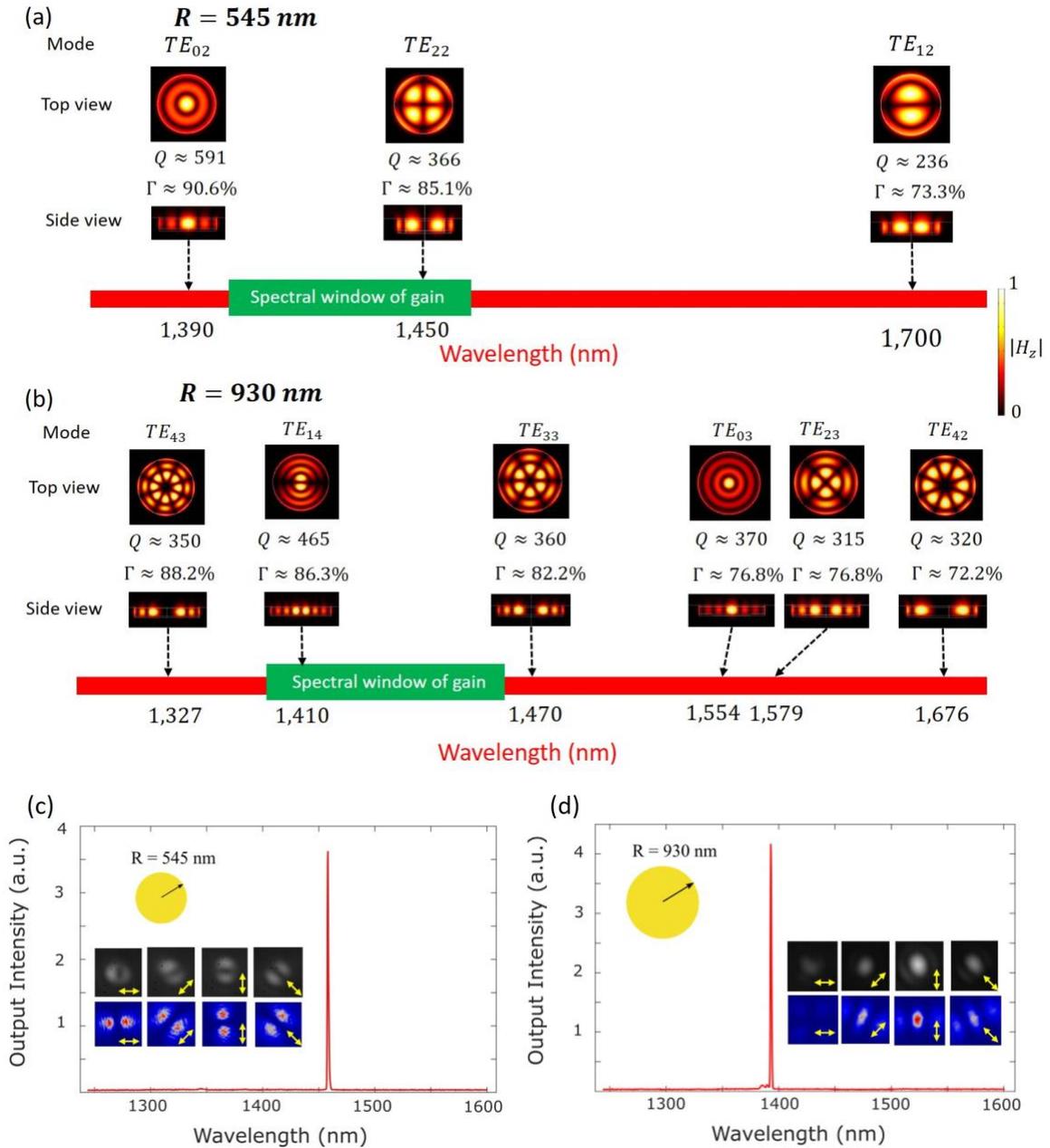

**Fig. S9| Resonance spectra and mode profiles associated with the highly confined modes of a single nanodisk laser.** Nanolaser elements used in (a) FM ($R = 545\ nm$) and (b) AF ($R = 930\ nm$) experiments, at a temperature of 225 K. In each case, $Q$ is the quality factor and $\Gamma$ is the gain confinement factor. These results explain the single-mode lasing of the $TE_{22}$ mode (case (a)) and $TE_{14}$ mode (case (b)) in each scenario. Measured emitted spectra from single nanolasers corresponding to (a, b) are depicted in (c, d) respectively. The insets in (c, d) indicate simulated as well as measured diffraction intensities and polarization states.

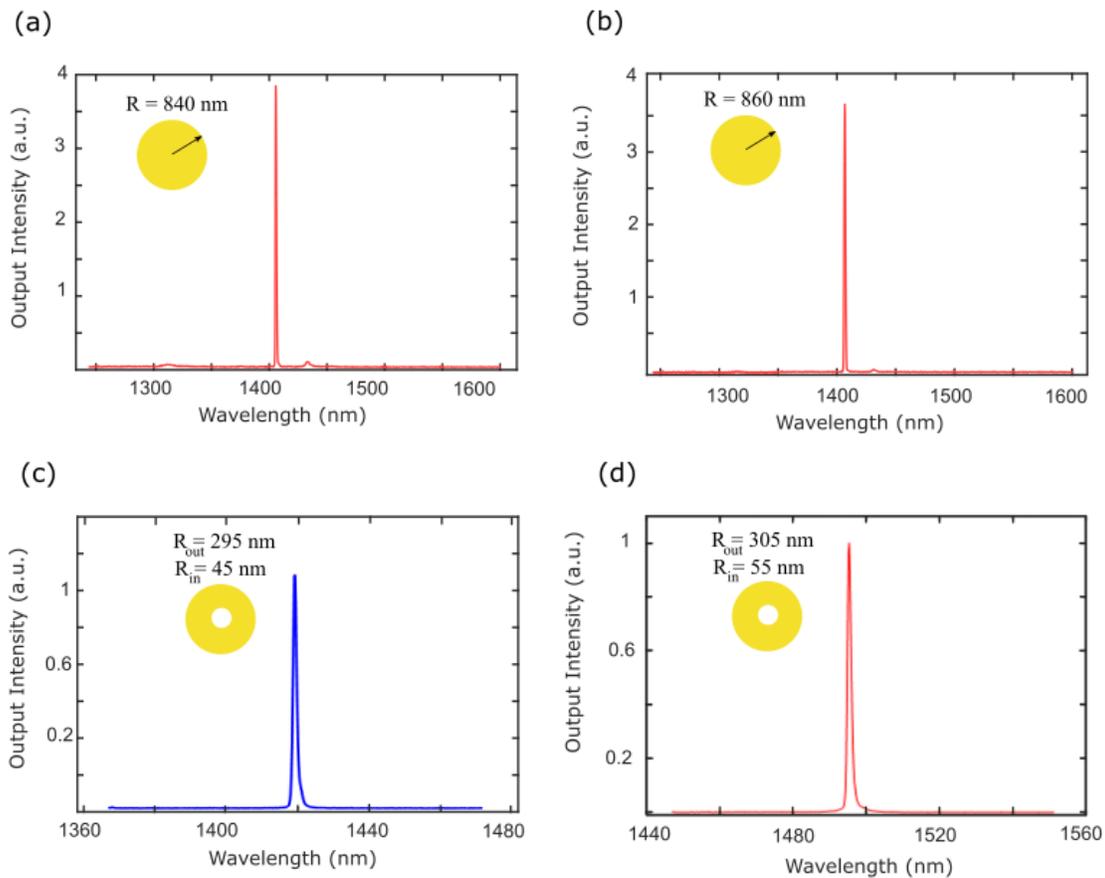

**Fig. S10| Single mode lasing action within various nanolaser elements.** Emitted light spectra from (a, b) nanodisks and (c, d) coaxial nanolasers with different sizes.

## Part 8. Lasing in the $TE_{0m}$ and $TM_{0m}$ modes

We considered using $TE_{0m}$ and $TM_{0m}$ modes of coaxial metallic nanolasers (Fig. S11) to realize Ising Hamiltonian in these platforms. Our preliminary simulations show that by using these modes it is possible to implement both FM and AF exchange interaction terms in the Ising Hamiltonian. This is evident from Eqs. (S-3) and (S-6), which in the case of $n = 0$ become:

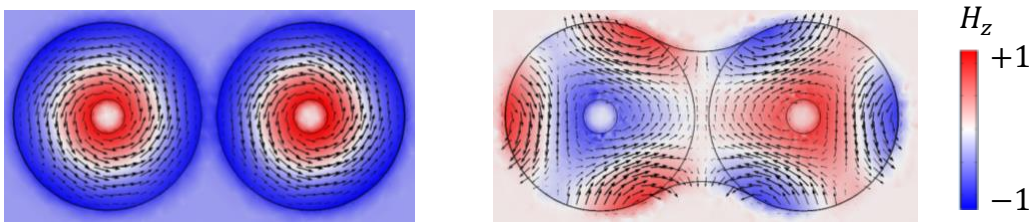

**Fig. S11| Lasing in the $TE_{0m}$ and $TM_{0m}$ modes of nanolasers.** Two coupled metallic coaxial nanolasers operating in a $TE_{01}$ mode, exhibiting an FM (left) and AF (right) exchange interaction between equivalent Ising pseudospins (spin up or down corresponds to clockwise/counterclockwise electric field vectors).

$$\mathcal{P}_{L,TE} \propto \mathcal{P}_1 - \mathcal{P}_{z,TE} \sum_{j=1}^{N} \cos\phi_j \cos\phi_{j+1} + \mathcal{P}_{\phi,TE} \sum_{j=1}^{N} \sin\phi_j \sin\phi_{j+1} \qquad \text{(S-12)}$$

$$\mathcal{P}_{L,TM} \propto \mathcal{P}_2 + \mathcal{P}_{\phi,TM} \sum_{j=1}^{N} \cos\phi_j \cos\phi_{j+1} \qquad \text{(S-13)}$$

In these scenarios, it can be directly seen from Eqs. (S-12, S-13) that the coupling between nearby lasers in this case is isotropic (independent of $\phi$, i.e. the local azimuthal coordinate of each element), and hence any anisotropic terms (e.g. the $H_0$ term) is absent in the associated equivalent Hamiltonian. Moreover, one can check that the ground state associated with the energy landscape functions in these equations satisfy $\phi_{j+1} = \phi_j + m\pi$. By utilizing this latter property, we intend to use such $TE_{0m}$ and $TM_{0m}$ modes to emulate the Ising Hamiltonian in our nanolaser platform.